\documentclass[journal]{IEEEtran}

\ifCLASSINFOpdf
\else

\fi

\usepackage[ruled,vlined]{algorithm2e}
\usepackage[hidelinks]{hyperref}
\usepackage[utf8]{inputenc}
\usepackage[small]{caption}
\usepackage{graphicx}{\graphicspath{figures}}
\usepackage{amsmath}
\usepackage{amsthm}
\usepackage{multicol}
\usepackage{multirow}
\usepackage{subcaption}
\usepackage{booktabs}
\usepackage{algpseudocode}

\begin{document}
%

\title{Single Image Internal Distribution Measurement Using Non-Local Variational Autoencoder}

%
%
%

\author{Yeahia Sarker,~\IEEEmembership{~Student Member,~IEEE},
        Abdullah-Al-Zubaer Imran,
        Md Hafiz Ahamed,~\IEEEmembership{~Graduated Member,~IEEE},
        Ripon K. Chakrabortty,
        Michael J. Ryan,~\IEEEmembership{~Senior Member,~IEEE},
        Sajal K. Das,~\IEEEmembership{~Member,~IEEE}

\thanks{Yeahia Sarker is with the Department of Mechatronics Engineering, Rajshahi University of Engineering \& Technology, Rajshahi 6204, Bangladesh. Email : yeahia.ruet@gmail.com}
\thanks{Abdullah-Al-Zubaer Imran is with Stanford University, Stanford, CA 94305, USA Email: aimran@stanford.edu}
\thanks{Md Hafiz Ahamed is with the Department of Mechatronics Engineering, Rajshahi University of Engineering \& Technology, Rajshahi 6204, Bangladesh. Email : hafiz@mte.ruet.ac.bd}
\thanks{Ripon K. Chakrabortty is with the School of Engineering and IT, University of New South Wales, Canberra 7916, Australia. Email : r.chakrabortty@unsw.edu.au}
\thanks{Michael J. Ryan is with the Capability Systems Centre, University of New South Wales, Canberra 7916, Australia. Email : m.ryan@unsw.edu.au}
\thanks{Sajal K. Das is with the Department of Mechatronics Engineering, Rajshahi University of Engineering \& Technology, Rajshahi 6204, Bangladesh. Email : das.k.sajal@gmail.com}
}

%
%

\markboth{
}{Sarker  \MakeLowercase{\textit{et al.}}: A Novel Learning Approach to Internal Distribution Measurement of Images using Non-Local Variational Autoencoder}
%



\maketitle

\begin{abstract}
\textbf{(This is a preprint)}. Deep learning-based super-resolution methods have shown great promise, especially for single image super-resolution (SISR) tasks. Despite the performance gain, these methods are limited due to their reliance on copious data for model training. In addition, supervised SISR solutions rely on local neighbourhood information focusing only on the feature learning processes for the reconstruction of low-dimensional images. Moreover, they fail to capitalize on global context due to their constrained receptive field. To combat these challenges, this paper proposes a novel image-specific solution, namely non-local variational autoencoder (\texttt{NLVAE}), to reconstruct a high-resolution (HR) image from a single low-resolution (LR) image without the need for any prior training. To harvest maximum details for various receptive regions and high-quality synthetic images, \texttt{NLVAE} is introduced as a self-supervised strategy that reconstructs high-resolution images using disentangled information from the non-local neighbourhood. Experimental results from seven benchmark datasets demonstrate the effectiveness of the \texttt{NLVAE} model. Moreover, our proposed model outperforms a number of baseline and state-of-the-art methods as confirmed through extensive qualitative and quantitative evaluations.

\end{abstract}
\section{Introduction}
Image super-resolution (SR) refers to the task of recovering a latent high-resolution (HR) image from a corresponding low-resolution (LR) image. This has been one of the most widely explored inverse problems in computer vision \cite{dian2017hyperspectral,liu2013infrared}. A LR image $I_{x}$ is assumed to be modeled as the output of the following degradation:
\begin{equation}
    I_{x}=\mathbf{D}(I_{y};\lambda).
\end{equation}
where $\mathbf{D}$ defines a  degradation mapping function, $I_{y}$ is the corresponding HR image, and $ \lambda$ denotes the degradation parameter. The higher the degradation, the harder becomes the task to reconstruct the HR image \cite{wang2020deep}. 
The image super-resolution problem is also explored in the fields of remote sensing \cite{lanaras2015hyperspectral}, surveillance imaging \cite{shamsolmoali2019deep}, and medical imaging \cite{mahapatra2019image}. While a number of approaches have been attempted to solve this problem, it remains ill-posed particularly since any specific LR image may correspond to croppings from multiple HR counterparts. 
Three types of solutions are usually provided to solve the SR problem; interpolation-based \cite{hung2011robust}, learning-based \cite{he2013beta,zeng2015coupled,yang2013fast} and reconstruction-based \cite{sun2010context,wang2014fast,fattal2007image}. Learning-based SR methods learn the non-linear mapping between HR and LR image using probabilistic generative models, random forest, linear or non-linear models, neighbor embedding \cite{chang2004super} and sparse regression \cite{kim2010single}. Interpolation-based methods utilize the adjacent pixels to calculate the interpolated pixels by using an interpolation kernel. Several types of existing interpolation-based models have been used to tackle the SR problem such as bicubic\cite{ruangsang2017efficient}, edge-directed estimation \cite{wang2013edge}, and auto-regressive models \cite{lu2017two}. Interpolation-based methods are very computationally efficient and relatively simple than other architectures. However, these methods suffer from low accuracy compared to other methods because of poor representation learning capacity \cite{shukla2020technical}. 

\begin{figure}[t!]
    \centering
     \subcaptionbox{HR Image}{
    \includegraphics[width = 0.37\linewidth, trim={0cm 8.5cm 7.5cm 0.88cm}, clip]{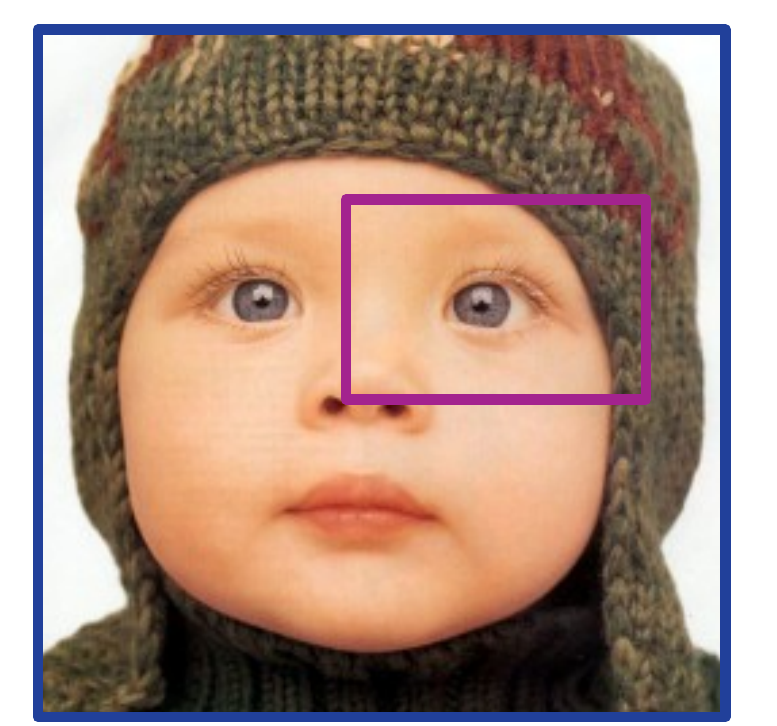}
    }
    \hspace{0.1cm}
    \centering
    \subcaptionbox{LR Image}{
    \includegraphics[width = 0.37\linewidth, trim={7.5cm 8.5cm 0cm 0.88cm}, clip]{introduction_picture.pdf}
    }
    \\[5pt]
    \centering
    \subcaptionbox{DIP}{
    \includegraphics[width = 0.37\linewidth, trim={0cm 0.88cm 7.5cm 9.25cm}, clip]{introduction_picture.pdf}
    }
    \hspace{0.1cm}
    \centering
    \subcaptionbox{NLVAE}{
    \includegraphics[width = 0.37\linewidth, trim={7.5cm 0.88cm 0cm 9.25cm}, clip]{introduction_picture.pdf}
    }
    \caption{Visual Comparison of Deep Image Prior (DIP) (untrained) \& NLVAE (untrained)}
    \label{fig:introduction_picture}
\end{figure}
\par
Learning-based solutions are widely used in the SR task. They can primarily be categorized into three types: Code-based\cite{freeman2002example,zhang2010partially,zeng2015coupled}, CNN-based \cite{kim2016accurate,gupta2018cnn} and regression-based \cite{zhang2012single}. In the past, learning-based solutions have shown great success for image super-resolution due to a robust feature learning capability \cite{ha2018deep,kim2016deeply}. The regression-based solutions are much faster than any other methods but, compared to other learning-based methods, produce blurry images and low peak-signal-noise ratio, due to poor representation learning  \cite{kim2010single,yang2013fast}. Learning-based solutions generally measure the similarity between LR \& HR images. A number of methods have been proposed to solve this problem but among them, CNN-based methods have superior performance because of their robust representation learning capabilities \cite{bhowmik2017training}. As a learning-based method, a hierarchical pyramid structure was developed using residual layers for image super-resolution \cite{lai2017deep}.
\par
Zhang \textit{et al.} \cite{zhang2018residual} introduced residual dense block in order to learn hierarchical feature maps, utilizing a bottleneck layer at the end of the residual dense layer. The local connection in the same block represents short-term memory and skip connections define long-term memory for representation learning. Lim \textit{et al.} \cite{lim2017enhanced} proposed Enhanced Deep Super-Resolution (EDSR) with a huge improvement in performance utilizing the residual structure. Despite the excellent performance, these methods utilize residual structure that is computationally resourceful. In \cite{hu2019channel}, a combination of channel-wise attention and spatial attention block was developed for single image super-resolution (SISR). Both of these blocks were again combined with residual creating robust SR methods. This method captures sufficient representations from feature space and suppresses irrelevant information.  As this approach combines two kinds of attention block with a stacked neural network, it consumes large amounts of memory and slows down the training process. 
\par
In \cite{mao2016image}, an autoencoder-based SISR method has been introduced with symmetric skip connections. Similarly, Tai \textit{et al.} \cite{tai2017image} featured a deep recursive residual network (DRRN)  utilizing memory block with residual representation learning. These solutions provide better results with the help of large scale architecture. Park \textit{et al.} \cite{park2018high} introduced a high dynamic range for super-resolution task that is very lightweight and easy to implement but decomposing the image causes loss of important information which results in low peak-signal noise ratio (PSNR) values.  
 
Many upsampling strategies have been observed in the literature. Efficient sub-pixel CNN (ESPCN) uses sub-pixel convolution for upsampling \cite{shi2016real} that stores channel information for extra points then reorganizes those points for HR reconstruction. Fast super-resolution CNN (FSRCNN) utilizes deconvolution operation to upsample \cite{dong2016accelerating}. Hua \textit{et al.} used a deconvolution operation for the upsampling process, featuring the arbitrary interpolation operator and subsequent convolution operator \cite{hua2019image}. It is to be noted that the deconvolution operation has two disadvantages: deconvolution is used at the end of the network, and the downsampling kernel is not known. Unknown input estimation consequently results in poor performance. To avoid these issues, we utilize linear upsampling of LR images so that we only focus on reconstruction quality rather than upsampling kernels.

\par
From a theoretical perspective, it can be deduced that the deeper neural architecture provides better result than shallow architecture \cite{montufar2014number}. Keeping this in mind, Kim \textit{et al.} \cite{kim2016accurate} first proposed a very deep architecture for SISR task. With 20 layers, VGGNet uses 3x kernels for all layers. Additionally, this method uses a high learning rate for faster convergence and utilizes gradient clipping to alleviate the gradient explosion problem. To learning short-term memory information, skip connections have been used in many tasks. Another work introduced recursive topology with parameter reduction using recursive convolution kernel \cite{tai2017image}. However, these settings are risky as self-supervised settings because a faster learning rate will provide shallow feature learning process, thus resulting in poor performance.

\par
Several reconstruction-based SISR methods have been introduced to solve the SISR problem, utilizing a shallow feature learning process \cite{lian2019fg,dou2020super}. KernelGAN \cite{bell2019blind}, consisting of a deep linear generator and a discriminator, supports blind SISR. A deep linear generator removes non-linear activation functions, but the overall loss function is not convex. The discriminator uses fully convolutional layers with no strides and pooling. Even though the overall structure means that the model converges faster, it is still difficult to obtain the global minimum. Our method utilizes the non-linear activation function in both encoder and decoder, making network learn more intuitive information than KernelGAN. Shaham \textit{et al.} \cite{shaham2019singan} proposed a multidisciplinary generative model capable of performing multiple computer vision tasks. To our knowledge, this work was the first attempt to use an unconditional generative model for the ZSSR task. It utilizes an adversarial network as a reconstruction-based method learning only abstract features from image patches. Due to the complexity of training an adversarial network, GAN models often suffer from convergence failure and mode collapse. Moreover, training adversarial training takes longer than discriminative models.

\par
To alleviate these problems, we have devised an image-specific architecture, called probabilistic non-local variational neural autoencoder (\texttt{NLVAE}), which can generate high-quality images with a robust pixel learning capability. Our generative solution is specifically designed for ZSSR, storing more disentangled and intuitive features and learning from low-dimensional space.

Our specific contributions can be summarized as follows:

\begin{itemize}
    \item An unconventional internal method has been introduced for the ZSSR task where only one one LR image and its corresponding HR image are required for the training process. The proposed method is comnpletely unsupervised and does not require any prior training. It establishes a new state-of-the-art (SOTA) which outperforms currently available methods.
    
    \item The proposed light-weight non-local feature extraction module harvests maximum representations from different receptive regions boosting the super-resolution performance.
    
    \item The proposed loss function aids to reconstruct high quality images by controlling the Lagrange multiplier and marginal value. 
    
\end{itemize}
The rest of the paper is organized as follows. Section II shows some works related to our proposed network structure. Section III describes the working principle of our method. In section IV, we provide quantitative and qualitative results using our model. Section V provides some ablation studies demonstrating the robustness of our network and section VI discusses the limitation of our strategy along with similarities and dissimilarities with other methods. Section VII provides concluding remarks.

\section{Related Work}
\textbf{Generative Models.} Generative models have been proven to reconstruct finer texture details and are able to generate more photo-realistic images than CNN-based methods. While shallow CNN-based SR methods provide detailed low-frequency information, GAN-based methods as generative models can discover high-frequency information. Super-Resolution GAN (SRGAN) \cite{ledig2017photo} makes use of perceptual loss as well as residual dense network generating high-resolution images. Wang \textit{et al.} \cite{wang2018esrgan} proposed residual-in-residual without batch normalization which produced HR images through adversarial training. Majdabad \textit{et al.} \cite{majdabadi2020capsule} attached a capsule network as a complex network with GAN for face super-resolution. In \cite{qiao2019image}, a conditional GAN has been introduced using ground-truth as a conditional variable for the discriminator. Similar to conditional GAN, conditional autoregressive generative models utilize maximum likelihood estimation depending on conditions. Based on these conditions, the generated HR images are reconstructed based on previous learned pixels \cite{van2016pixel,van2016conditional}. However, these generative models suffer from mode collapse and convergence failure problems \cite{goodfellow2016nips}. Moreover, these methods are computationally expensive and integration with self-supervised training is quite difficult to implement as GAN-based methods require more image data for training than learning-based methods \cite{takano2019srgan,wang2018esrgan}.
\par
\textbf{Non-Local Networks.} Non-local networks usually comprise an attention module with non-local blocks. Wang \textit{et al.} \cite{wang2019deformable} proposed a deformable non-local attention module for video super-resolution. In \cite{liu2018non}, a non-local recurrent model was introduced for SISR task, which can learn deep feature correlation among neighbourhood locations of patches. Zhang \textit{et al.} \cite{zhang2018residual} featured residual network with non-local attention units for image super-resolution. Another work presents a cross-scale non-local attention module for learning intrinsic feature correlations of images \cite{mei2020image}. 
\par

\begin{figure}[t!]
    \centering
    \includegraphics[width = 0.48\textwidth]{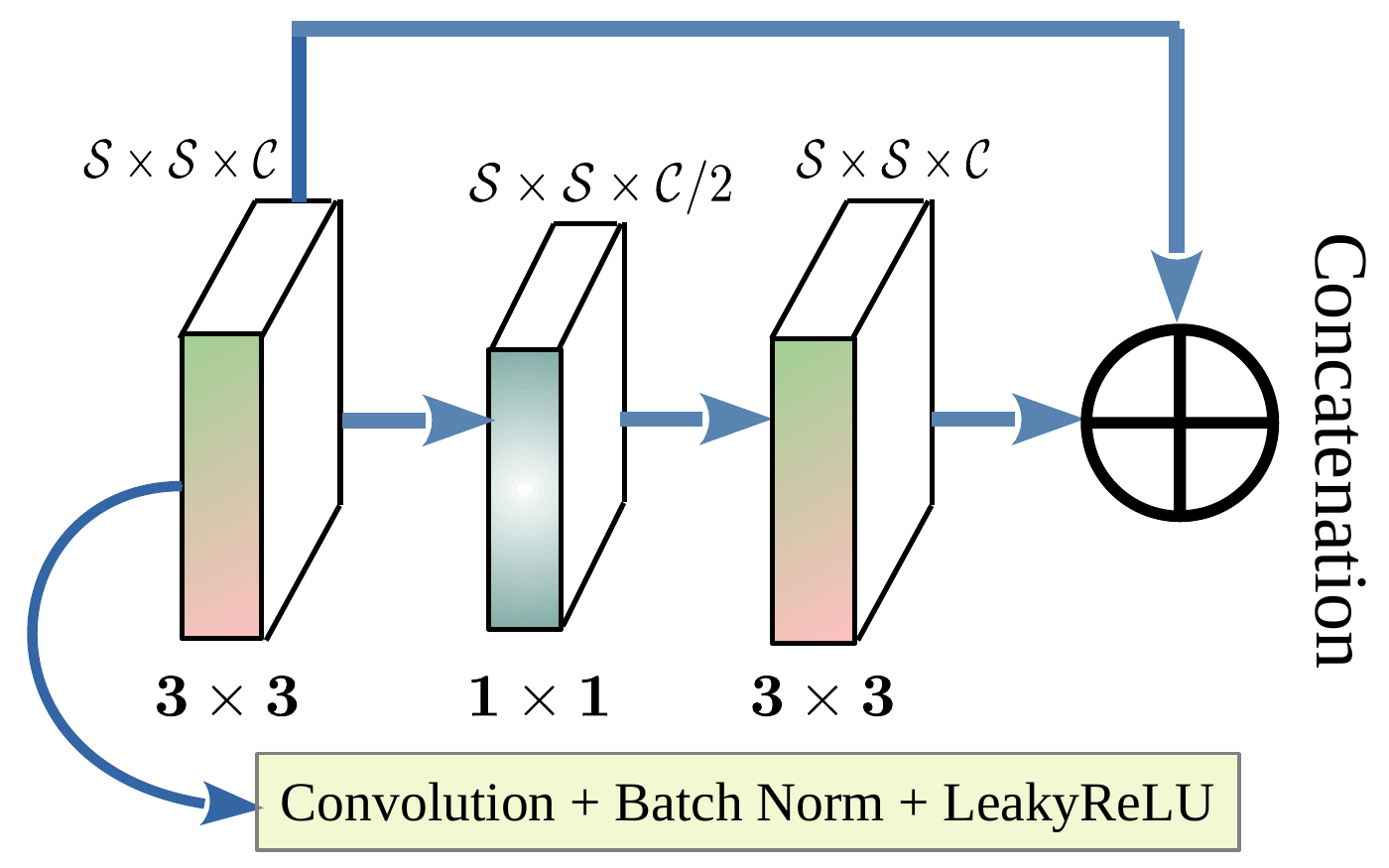}
    \caption{Overview of non-local block used in our \texttt{NLVAE} network. The non-local block is composed of 3$\times$ kernels and 1$\times$ kernels. The initial convolution kernel is concatenated with the last feature transform to learn relative positional features. $S\times{S}$ defines the spatial size of the feature and the channel information is denoted as $C$}
    \label{fig:non_local}
\end{figure}

\textbf{Zero-Shot Super-Resolution Methods.} Shocher \textit{et al.} \cite{shocher2018zero} introduced the term "ZSSR," presenting a shallow CNN model to learn the probability distribution of the LR and HR images. The major disadvantage of this network is that it extracts local features utilizing a simple CNN architecture and a shallow CNN model, which also results in poor performance. Another internal method proposed in \cite{ulyanov2018deep}, introducing Deep Image Prior (DIP) to build a bridge between a CNN and convolutional sparse coding. The solution takes the neural network as the output of the reconstruction and random input signals. This is the first approach which creates a bridge between a code-based method and a learning-based method for ZSSR. Untrained DIP basically focuses on the smaller receptive fields for intuitive neural representation, but loses more context as feature extraction is limited to the smaller regions. Fig. ~\ref{fig:introduction_picture} depicts that DIP method shows very weak structural information compared to \texttt{NLVAE}. Due to a weak feature extraction process, this method suffer from low accuracy in terms of performance metrics \cite{bengio2013representation}.

\begin{figure*}[t!]
    \centering
    \includegraphics[width = 0.95\linewidth]{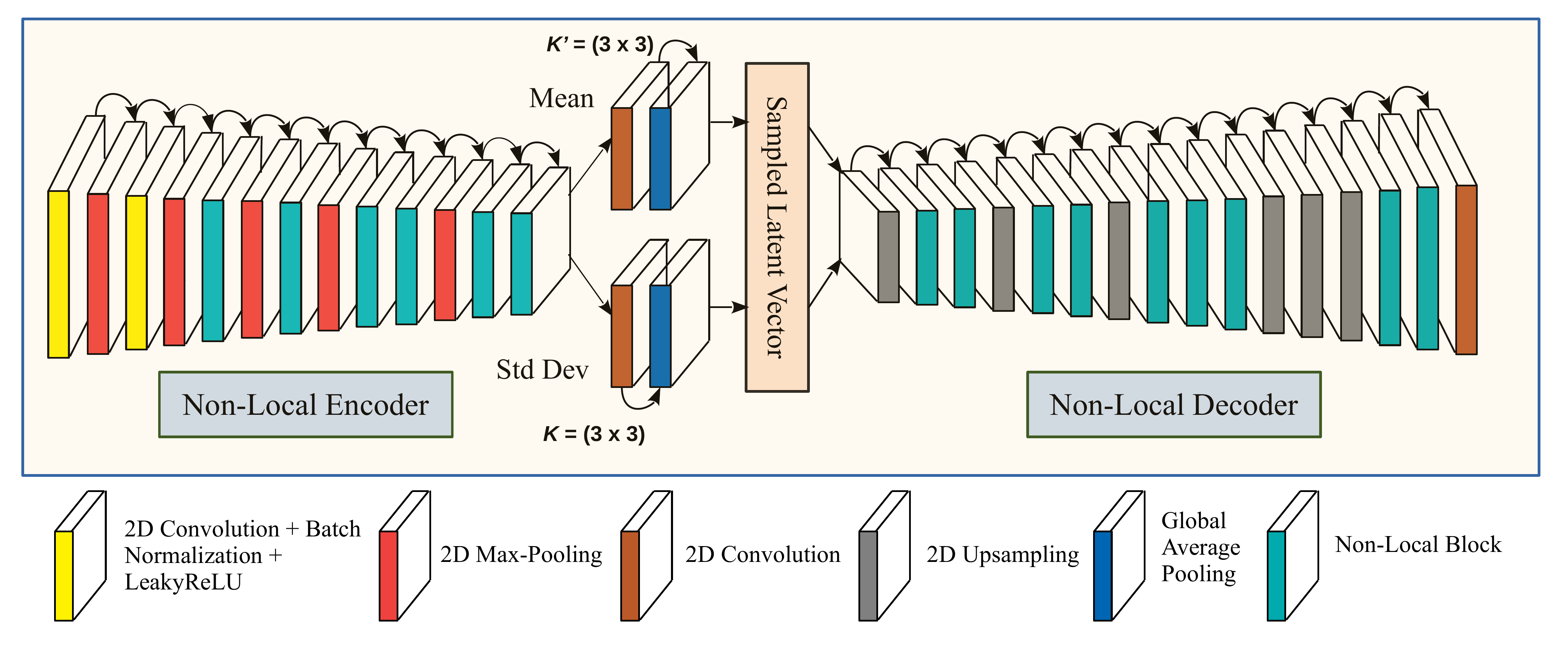}
    \caption{Network structure of the proposed non-local variational autoencoder (\texttt{NLVAE}) model. Probabilistic encoder-deocder are composed of non-local units and various convolution and upsampling layers. The reconstruction quality is controlled by the operator $\beta$. Global avg pooling are used to calculate the mean and variance leveraging global structural details during the reconstruction process.}
    \label{fig:nlvae}
\end{figure*}

\section{Network Structure}
In this section, we demonstrate the structure of our proposed non-local block in the neural encoder and decoder. We also show the measurement of posterior distribution and loss function and we provide a full analysis of how the Lagrange multiplier controls the reconstruction quality of generated image.

\subsection{Non-Local Encoder-decoder}
As shown in Fig.~\ref{fig:nlvae}, our proposed \texttt{NLVAE} model consists of an encoder and a decoder consisting of non-local convolution blocks. Fig.~\ref{fig:non_local} depicts the overview of non-local blocks. The non-local block utilized in \texttt{NLVAE} can exploit spatial correlation between neighbourhood and locations. To design a computationally effective spectral correlation module, we have overlooked residual structure. Each non-local block is composed of convolution blocks and each convolution block comprises of 2D convolution, point-wise convolution and batch normalization \cite{ioffe2015batch} followed by Leaky-ReLU activation function. The encoder encodes the input image $x$ into the latent representation $z$ ( $z= \psi (x)$) and the decoder reconstructs the representation back to its approximate original data. We assume that the low-resolution image is an input vector denoted by $x$ and $z$ denotes the latent representation. The latent variables are controlled by a Gaussian distribution along with a diagonal covariance matrix. The latent space dimension is denoted by $J$. The output of the non-local convolutional encoder comprises a mean $\mu$ and a log of variances $log(\sigma)$. Through the reparameterization trick, a noise vector $(\in)$ is obtained from the latent space \cite{kingma2013auto}. The goal of the \texttt{NLVAE} model is the ability to produce a high-resolution reconstructed image from the low-resolution image, exploiting the relationship between the input vector and prior distribution $p(z)$.

\subsection{Posterior Distribution}
\vspace{-0.07cm}
We denote the low-resolution input data distribution as $ x ~ p_{d}x$ and the high-resolution reconstructed data distribution as $\bar{x} ~ p(x) $. The encoded and decoded data distributions are represented as $q_{\phi}(z|x)$ and $p_{\theta}(z|x)$ respectively, where $\phi$ and $\theta$ are the variables of the encoder and decoder networks. The $q(x)$ tries to approximate the output prior $p(z)$. Centered isotropic multivariate Gaussian $N(0,I)$ was chosen as the prior $p(z)$ over the latent variables \cite{durrieu2012lower}. The inference model is designed to output two individual variables $\mu$ and $\sigma$, and thus the posterior $q_{\phi}(z|x) = N(z;\mu,\sigma_{2})$. With this setting, to get the desired prior distribution, the non-local convolutional encoder and decoder are trained to optimize the reconstruction error (that is, the mean squared error). The loss function tries to approximate between each patch of LR and HR over a fake minibatch $x_{n}$ where $n$ is the size of the fake minibatch. The total number of data points is denoted as $N$.

\begin{algorithm}[t!]
\caption{Training \texttt{NLVAE} model}
\label{alg:nlvae}
\SetAlgoLined
\textbf{Input : }Initialize network parameters

 \While{not converged}{
 
  $\mathbf{X} =$ psuedo labels of single $L_{r}$ image
  
  $Z_{p} \leftarrow$ Distribution from prior $ \textbf{N}(0,I)$ 
  
  $Z_{p} \leftarrow \mathbf{Non-Local Encoder}\big(X \big)$ 
  
  $X_{r} \leftarrow Non-Local Decoder\big(Z\big)$ 
  
  $X_{p} \leftarrow Non-Local Decoder\big(Z_{p}\big)$
  
  $L_{KL} \leftarrow L_{KL}\big(X_{r}, X_{p}\big)$
  
  $\phi_{E} \leftarrow \phi_{E} - \eta \nabla_{\phi e}(L_{R}+\beta L_{AE}\big)$
  
  (Updating Adam for $\phi_{E}$)
  }

\end{algorithm}

The input of the decoder is sampled from $\mathbf{N}(z;\mu,\sigma_{2})$ using a reparameterization trick $z = \mu + \sigma \odot \epsilon$ where $\epsilon = \mathbf{N} (0,I)$. The aggregated posterior distribution is defined as $z = q(z)$:
\begin{equation}
\label{eqn:posterior}
q(\mathbf {z}) = \int _{\mathbf {x}} q_{\phi }(\mathbf {z}|\mathbf {x})p_{d}(\mathbf {x}) d\mathbf {x}.
\end{equation}

\subsection{Loss Function}

It is useful to prepare the low-resolution input image by clustering in latent space, eradicating the noise $L_{R}$ loss is summed over the data points and the average of the fake minibatch is calculated. Thus, it provides more weight to the reconstruction error helping reduce potential model collapse. 
\begin{equation}
\label{eqn:reconstruction_loss}
{L}_{R}({\phi }, {\theta };{x}_{M}, {\epsilon }) = \frac {1}{M} \sum _{i=1}^{M} \sum _{j=1}^{N}({x}_{i,j} - \hat {{x}}_{i,j})^{2}.
\end{equation} 

To obtain the desired prior distribution, $KL$ divergence is utilized on the encoded variable to measure the probability distance of LR and HR images. $KL$ divergence is calculated over the fake minibatch as
\begin{equation}
\label{eqn:kl_loss}
{L}_{KL}( {\phi }; {x}_{M}) \\= \frac {1}{M} \sum _{i=1}^{M}\sum _{j=1}^{J} \left ({1 + log(\sigma _{i,j} ^{2}) - \mu _{i,j}^{2} - \sigma _{i,j} ^{2} }\right) .
\end{equation}

Therefore, the total loss is calculated as 
\begin{equation}
\label{eqn:total_loss}
{L}_{BETA}({\phi },{\theta };{x}_{M},{\epsilon }) = {L}_{R} + \beta{L}_{KL} + \alpha
\end{equation}

where $\beta$ denotes the Lagrangian multiplier and $\alpha$ denotes the marginal value. As the negation of $L_{BETA}$ is the lower bound of the Lagrangian, minimization of the loss is equivalent to maximization of the Lagrangian, which is useful for our initial optimization problem. The $\alpha$ controls the quality of image reconstruction as an aid to the objective function. For $\beta=1$, the working principle is the same as traditional VAE. When $\beta > 1$, it applies a stronger constraint on the latent bottleneck and limits the representation capacity of $z$ \cite{chen2018isolating}. Maintaining the disentanglement is the most effective representation for some of the conditionally independent generative factors \cite{mathieu2019disentangling}.

\subsection{Lagrangian multiplier variation for different upscaling}
The addition of $\beta $ in VAE provides more disentangled information and sharp gradients compared to traditional VAE \cite{higgins2016beta}. A higher value of $\beta$ provides more efficient encoded latent vectors and further encourages disentanglement. However, too large $\beta$ may lead to poorer reconstruction quality as it creates a trade-off with the extent of disentanglement. The reconstruction loss ensures the network captures useful information while forming the latent distribution. An increase in the number of latent variables reduces image quality, thus through empirical evaluation, we selected a different Lagrangian multiplier for different upscaling. For our experimental settings, we have selected 150, 200, and 300 for 3 $\times$, 4 $\times$, and 8 $\times$ upscaling factor respectively. 

\subsection{Computational Efficiency of Non-local Block}
In this subsection, the computational efficiency of the non-local block is briefly explained. The point-wise convolution is the core of non-local block for calibrating spatial information. It also serves as the channel reduction technique in this network. The weights of the point-wise convolution can be calculated as:
\begin{align}
\label{eqn:point-wise}
    W_{PC} = K \times K \times N \times P.
\end{align}
For this operation, $K = 1$. Then Equation~\ref{eqn:point-wise} becomes:
\begin{align}
        \label{eqn:point_weights}
        W_{PC} =  N \times P.
\end{align}
And the corresponding number of operations is therefore:
\begin{align}
    \label{eqn:point_operation}
    O_{PC} = M \times M \times K \times K \times N \times P \\ \nonumber
    = M \times M \times N \times P. &&
\end{align}

For the standard convolution operation, the number of weights will be:
\begin{align}
    \label{eqn:std_conv_weights}
    W_{SC} = K \times K \times N \times P.
\end{align}
And the corresponding number of operation is:
\begin{align}
    \label{eqn:std_conv_operation}
     O_{SC} = M \times M \times K \times K \times N \times P.
\end{align}
Now, the reduction factors of weights and operations can be defined as:
\begin{align}
    \label{eqn:reduction_weights}
    F_{W} = \frac{W_{PC}}{W_{SC}}.
\end{align}
\begin{align}
    \label{eqn:reduction_operation}
    F_{O} = \frac{O_{PC}}{O_{SC}}.
\end{align}
From the reduction factors of weights and operation, we can observe the reduction in computational cost due to the use of point-wise convolutions.

\begin{table*}[ht!]
\setlength{\tabcolsep}{4pt}
\centering
\caption{Benchmark results for SISR methods. Best results are in bold. All the methods are trained on DIV2K datasets.}
\medskip
\label{tab:performance}
\begin{tabular}{@{} cc l c cc c cc c cc c cc c cc @{}}
\toprule
\multirow{2}{*}{Scale} & 
\phantom{a} & 
\multirow{2}{*}{Method} & 
\phantom{a} &  
\multicolumn{2}{c}{Set5} & 
\phantom{a} & 
\multicolumn{2}{c}{Set14} & 
\phantom{a} & 
\multicolumn{2}{c}{BSDS100} & 
\phantom{a} & 
\multicolumn{2}{c}{Urban100} & 
\phantom{a} & 
\multicolumn{2}{c}{Manga109} \\
\cmidrule{5-6}
\cmidrule{8-9}
\cmidrule{11-12}
\cmidrule{14-15}
\cmidrule{17-18}

&& && PSNR & SSIM && 
PSNR  & SSIM &&
PSNR  & SSIM &&
PSNR  & SSIM &&
PSNR  & SSIM \\
\midrule

\multirow{10}{*}{\rotatebox[]{45}{3$\times$}} && 
Bicubic &&
30.40 & 0.8684 &&
27.55 & 0.7743 &&
27.19 & 0.7388 &&
24.45 & 0.7358 &&
26.95 & 0.8558 \\

&& 
A+ \cite{timofte2014a+} &&
32.51  & 0.9080 &&
29.10  & 0.8202 && 
28.21  & 0.7829 && 
25.86  & 0.7891 &&
29.90  & 0.9101 \\

&&
SRCNN \cite{dong2015image} &&
32.75 & 0.9090 &&
29.30 & 0.8215 &&
28.28 & 0.7832 &&
25.87 & 0.7888 &&
30.56 & 0.9124 \\

&&
FSRCNN \cite{dong2016accelerating} &&
33.17 & 0.9141 && 
29.39 & 0.824 && 
28.59 & 0.7940 && 
26.43 & 0.8075 &&
31.05 & 0.9189 \\

&&
VDSR \cite{kim2016accurate} &&
33.67 & 0.9212 &&
29.78 & 0.8318 &&
28.83 & 0.7982 &&
27.14 & 0.8280 && 
32.07 & 0.9337 \\

&&
LapSRN \cite{lai2017deep} &&
33.82 & 0.9227 &&
29.84 & 0.8322 &&
28.82 & 0.7982 &&
27.07 & 0.8270 &&
32.21 & 0.9342 \\ 

&&
MemNet \cite{tai2017memnet} &&
34.09 & 0.9248 &&
30.00 & 0.8350 &&
28.95 & 0.8001 &&
27.53 & 0.8270 &&
32.58 & 0.9382 \\

&&
SRGAN \cite{ledig2017photo}  &&
33.73 & 0.9102 &&
29.58 & 0.8215 &&
28.62 & 0.7790 &&
26.04 & 0.8168 &&
31.56 & 0.9187 \\

&&
NLVAE (Proposed) &&
\textbf{34.10} & \textbf{0.9270} &&
\textbf{30.81} & \textbf{0.8398} &&
\textbf{29.05} & \textbf{0.7805} &&
\textbf{28.07} & \textbf{84.02} &&
\textbf{33.19} & \textbf{0.9437} \\ 
\midrule

\multirow{10}{*}{\rotatebox[]{45}{4$\times$}} &&
Bicubic && 
28.43 & 0.8109 &&
26.00 & 0.7026 &&
25.95 & 0.6698 &&
23.13 & 0.6598 &&
24.89 & 0.7865 \\

&&
A+ \cite{timofte2014a+} &&
30.25 & 0.8601 &&
27.21 & 0.7503 &&
26.65 & 0.7103 &&
24.19 & 0.7198 &&
27.08 & 0.8519 \\

&&
SRCNN \cite{dong2015image} &&
30.48 & 0.8628 && 
27.50 & 0.7513 &&
26.90 & 0.7114 &&
24.52 & 0.7221 &&
27.60 & 0.8583 \\

&&
FSRCNN \cite{dong2016accelerating} &&
30.72 & 0.8658 &&
27.60 & 0.7538 &&
26.95 & 0.7138 &&
24.62 & 0.7280 &&
27.86 & 0.8602 \\

&&
VDSR \cite{kim2016accurate} &&
31.35 & 0.8838 &&
28.02 & 0.7682 &&
27.29 & 0.7165 &&
25.18 & 0.7530 &&
28.87 & 0.8862 \\

&&
LapSRN \cite{lai2017deep} &&
31.54 & 0.8860 &&
28.16 & 0.7724 &&
27.32 & 0.7161 &&
25.21 & 0.7558 &&
29.09 & 0.8890 \\

&&
MemNet \cite{tai2017memnet} &&
31.76 & 0.8893 &&
28.26 & 0.7726 &&
27.42 & 0.7280 &&
25.50 & 0.7628 &&
29.64 & 0.8938 \\

&&
SRGAN \cite{ledig2017photo} &&
29.37 & 0.8471 &&
26.01 & 0.7396 &&
25.13 & 0.6645 &&
24.35 & 0.7331 &&
28.39 & 0.8603 \\

&&
ESRGAN \cite{wang2018esrgan} &&
30.47 & 0.8512 &&
26.28 & 0.6987 &&
25.32 & 0.6519 &&
24.36 & 0.7337 &&
28.44 & 0.8609 \\

&&
NLVAE (Proposed) &&
\textbf{31.96} & \textbf{0.8903} && 
\textbf{28.67} & \textbf{0.7776} && 
\textbf{27.86} & \textbf{0.7367} && 
\textbf{25.88} & \textbf{0.7751} &&
\textbf{30.11} & \textbf{0.8945} \\
\midrule

\multirow{10}{*}{\rotatebox[]{45}{8$\times$}} && 
Bicubic && 
24.42  & 0.6580 &&
23.10 & 0.5660 &&
23.65 & 0.5483 &&
20.74 & 0.5160 &&
21.55 & 0.6509 \\

&&
A+ \cite{timofte2014a+} &&
25.21 & 0.6875 &&
23.48 & 0.5889 &&
23.97 & 0.5605 &&
21.02 & 0.5403 &&
22.11 & 0.6813 \\

&&
SRCNN \cite{dong2015image} &&
25.33 & 0.6900 &&
23.76 & 0.5910 &&
24.13 & 0.5659 &&
21.29 & 0.5438 &&
22.40 & 0.6846 \\ 

&&
FSRCNN \cite{dong2016accelerating} &&
20.13 & 0.5520 && 
19.75 & 0.4820 &&
24.21 & 0.5672 &&
21.32 & 0.5379 &&
22.39 & 0.6730 \\

&&
VDSR \cite{kim2016accurate} &&
25.95 & 0.7242 && 
24.26 & 0.6140 &&
24.37 & 0.5767 &&
21.65 & 0.5704 &&
23.16 & 0.7230 \\ 

&&
LapSRN \cite{lai2017deep} &&
26.14 & 0.7384 &&
24.35 & 0.6200 &&
24.53 & 0.5865 &&
21.81 & 0.5805 &&
23.39 & 0.7533 \\

&&
MemNet \cite{tai2017memnet} &&
26.16 & 0.7414 &&
24.38 & 0.6199 &&
24.59 & 0.5843 &&
21.88 & 0.5824 &&
23.56 & 0.7386 \\ 

&&
SRGAN \cite{ledig2017photo} &&
25.88 & 0.7069 &&
24.02 & 0.6015 &&
24.41 & 0.5786 &&
21.68 & 0.5614 &&
24.61 & 0.7864 \\

&&
ESRGAN \cite{wang2018esrgan} &&
26.30 & 0.7551 &&
24.07 & 0.6011 &&
24.64 & 0.5850 &&
22.57 & 0.6279 &&
24.75 & 0.7872 \\

&&
NLVAE (Proposed) &&
\textbf{27.23} & \textbf{0.7860} && 
\textbf{25.32} & \textbf{0.6469} && 
\textbf{25.31} & \textbf{0.5983} &&
\textbf{22.97} & \textbf{0.6353} &&
\textbf{25.12} & \textbf{0.8013} \\
\bottomrule

\end{tabular}
\end{table*}

For the standard convolution operation, the number of weights will be
\begin{align}
    \label{eqn:std_weights}
    W_{SC} = K \times K \times N \times P.
\end{align}
And the corresponding number of operation is,
\begin{align}
    \label{eqn:std_operation}
     O_{SC} = M \times M \times K \times K \times N \times P.
\end{align}
Now, the reduction factors of weights and operations can be defined as
\begin{align}
    \label{eqn:reduction_weights}
    F_{W} = \frac{W_{PC}}{W_{SC}}.
\end{align}
\begin{align}
    \label{eqn:reduction_operation}
    F_{O} = \frac{O_{PC}}{O_{SC}}.
\end{align}
From the reduction factors of weights and operation, we can observe the reduction in computational cost due to the use of point-wise convolutions.

\section{Experimental Evaluations}

\subsection{Datasets}
We have evaluated our proposed \texttt{NLVAE} model against seven different datasets---Set5 \cite{bevilacqua2012low}, Set14 \cite{zeyde2010single}, BSD100 \cite{martin2001database}, Manga109 \cite{matsui2017sketch}, Urban100 \cite{huang2015single}, General100 \cite{dong2016accelerating}, and T19 \cite{yang2010image}. In the qualitative analysis, we have used the General100, Set14, Set5, and T91 datasets, while the quantitative analyses are performed using the Set5, Set14, BSD100, Urban100, and Manga109 datasets. We have compared our model against a number of baseline and SOTA models, reporting PSNR and SSIM metrics.

\subsection{Implementation Details \& Training Settings}

We make use of the TensorFlow framework with Python for all the experiments. The experiments are implemented on a a Nvidia GeForce GTX Titan X and Intel Xeon CPU at 2.40 GHz machine. All the images are resized to $256 \times 256$. The Adam optimizer \cite{kingma2014adam} is utilized with $\beta_{1}$ = 0.9, $\beta_{2}$ = 0.999, and $\sigma$ = $10^{-8}$. Pesudo labels are created for training purposes as there exists only one single image. The model is trained till 2000 epochs. We use the $L2$ loss function for our solution. For our settings, the hyperparameters are selected empirically. We perform the experiments for three different scaling factors---$3\times$, $4\times$, and $8\times$. The value of $\beta$ is set to 500 for all the experiments.
\par
As the proposed method utilizes self-training strategy, it takes both $L_R$ and $H_R$ single image as input for training pipeline. Then, it tries to learn the relationship leveraging non-local attention blocks. Finally, the self-training model tries to generate a single $H_R$ image from the pre-trained weights. It is to be noted that the performance metric calculation is the mean of all single generated images for each datasets.

\subsection{Results}

\begin{figure*}[ht!]
\centering
 \resizebox{0.9\linewidth}{!}{%
  \begin{tabular}{c c c c c}
  & {\Large HR Image} & {\Large SRGAN} & {\Large ESRGAN} & {\Large NLVAE} \\
  \noalign{\smallskip}
  {\Large \rotatebox{90} {\hspace{1.5cm}{tt17}}} & 
  \includegraphics[width=0.24\linewidth, trim={1.25cm 15.5cm 23cm 1.1cm}, clip]{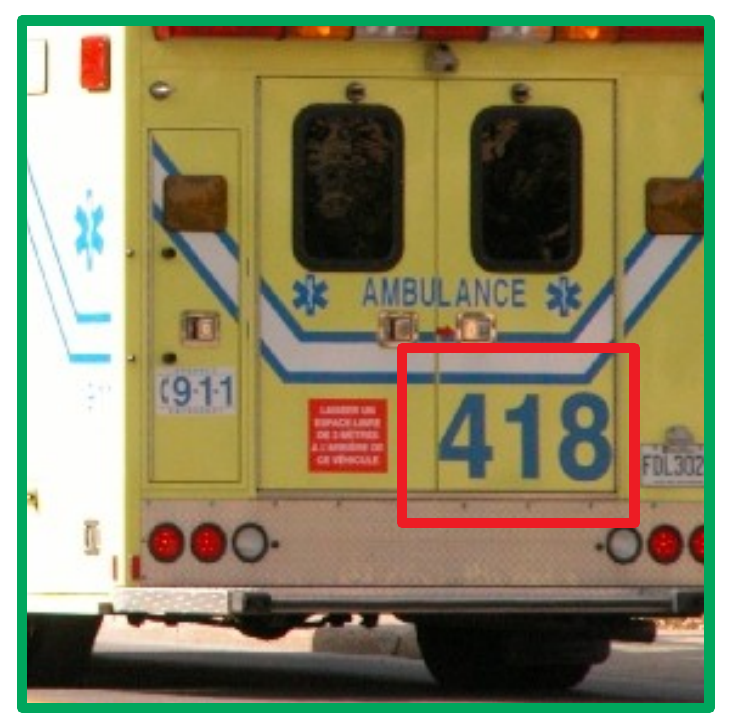} &
  \includegraphics[width=0.24\linewidth, trim={9cm 15.5cm 15.25cm 1.1cm}, clip]{generative_model.pdf} &
  \includegraphics[width=0.24\linewidth, trim={16.25cm 15.5cm 8cm 1.1cm}, clip]{generative_model.pdf} &
  \includegraphics[width=0.24\linewidth, trim={24cm 15.5cm 0.25cm 1.1cm}, clip]{generative_model.pdf} 
  \\
  \noalign{\smallskip}
  
    {\Large \rotatebox{90} {\hspace{1cm}{butterfly}}} & 
    \includegraphics[width=0.24\linewidth, trim={1.25cm 7.9cm 23cm 8.7cm}, clip]{generative_model.pdf} &
    \includegraphics[width=0.24\linewidth, trim={9cm 7.9cm 15.25cm 8.7cm}, clip]{generative_model.pdf} &
    \includegraphics[width=0.24\linewidth, trim={16.25cm 7.9cm 8cm 8.7cm}, clip]{generative_model.pdf} &
    \includegraphics[width=0.24\linewidth, trim={24cm 7.9cm 0.25cm 8.7cm}, clip]{generative_model.pdf} 
     \\
     \noalign{\smallskip}
     
     {\Large \rotatebox{90} {\hspace{1.4cm}{img\_087}}} & 
    \includegraphics[width=0.24\linewidth, trim={1.25cm 0.1cm 23cm 16.5cm}, clip]{generative_model.pdf} &
    \includegraphics[width=0.24\linewidth, trim={9cm 0.1cm 15.25cm 16.5cm}, clip]{generative_model.pdf} &
    \includegraphics[width=0.24\linewidth, trim={16.25cm 0.1cm 8cm 16.5cm}, clip]{generative_model.pdf} &
    \includegraphics[width=0.24\linewidth, trim={24cm 0.1cm 0.25cm 16.5cm}, clip]{generative_model.pdf} 
    \\
 \end{tabular}
  }
  \caption{Visual comparison of reconstruction-based methods on 'tt17.png' from T91 dataset and 'butterfly.png' from Set5 dataset and 'img\_087.png' Urban100}
  \label{fig:reconstruction}
  \end{figure*}

\begin{figure*}[ht!]
\centering
 \resizebox{0.9\linewidth}{!}{%
  \begin{tabular}{c c c c c}
  & {\Large HR Image} & {\Large VDSR} & {\Large LapSRN} & {\Large NLVAE} \\
 \noalign{\smallskip}
  {\Large \rotatebox{90}{\hspace{1cm} {im\_078}}} &
  \includegraphics[width=0.24\linewidth, trim={1.25cm 7.9cm 23cm 1.25cm}, clip]{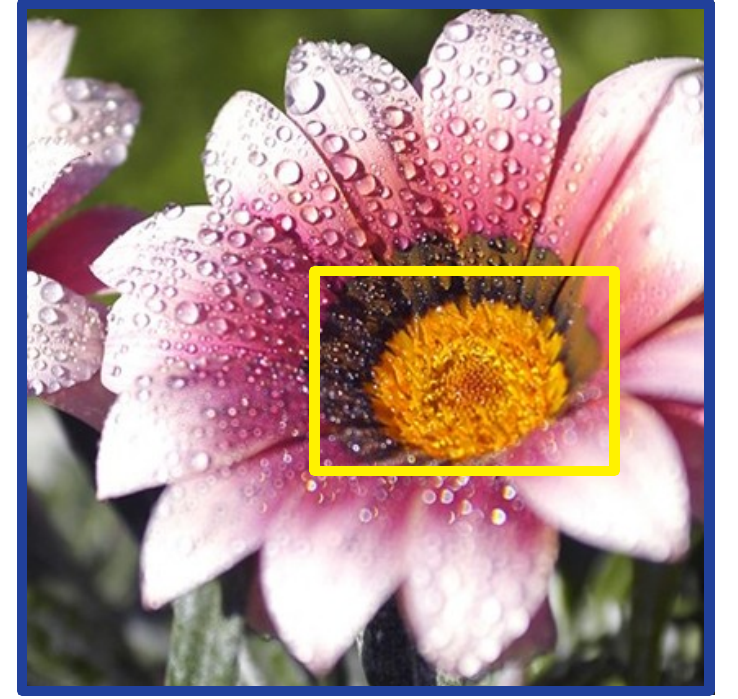} &
    \includegraphics[width=0.24\linewidth, trim={9cm 7.9cm 15.25cm 1.25cm}, clip]{learning.pdf} &
      \includegraphics[width=0.24\linewidth, trim={16.25cm 7.9cm 8cm 1.25cm}, clip]{learning.pdf} &
      \includegraphics[width=0.24\linewidth, trim={24cm 7.9cm 0.25cm 1.25cm}, clip]{learning.pdf} 
      \\
           \noalign{\smallskip}

        {\Large \rotatebox{90}{\hspace{1.4cm}{pepper}}} &
  \includegraphics[width=0.24\linewidth, trim={1.25cm 0.25cm 23cm 8.75cm}, clip]{learning.pdf} &
    \includegraphics[width=0.24\linewidth, trim={9cm 0.25cm 15.25cm 8.75cm}, clip]{learning.pdf} &
      \includegraphics[width=0.24\linewidth, trim={16.25cm 0.25cm 8cm 8.75cm}, clip]{learning.pdf} &
      \includegraphics[width=0.24\linewidth, trim={24cm 0.25cm 0.25cm 8.75cm}, clip]{learning.pdf} 
 \end{tabular}
  }
    \caption{Visual comparison of learning-based methods on 'im\_078.png' from General100 dataset and 'pepper.png' from Set14 dataset.}
    \label{fig:learning}
  \end{figure*}

\begin{figure*}[ht!]
    \centering
    \resizebox{0.95\linewidth}{!}{
    \subcaptionbox{L1 \& L2 loss functions, and various optimizers with respect to epochs.}{
    \includegraphics[width = 0.45\linewidth]{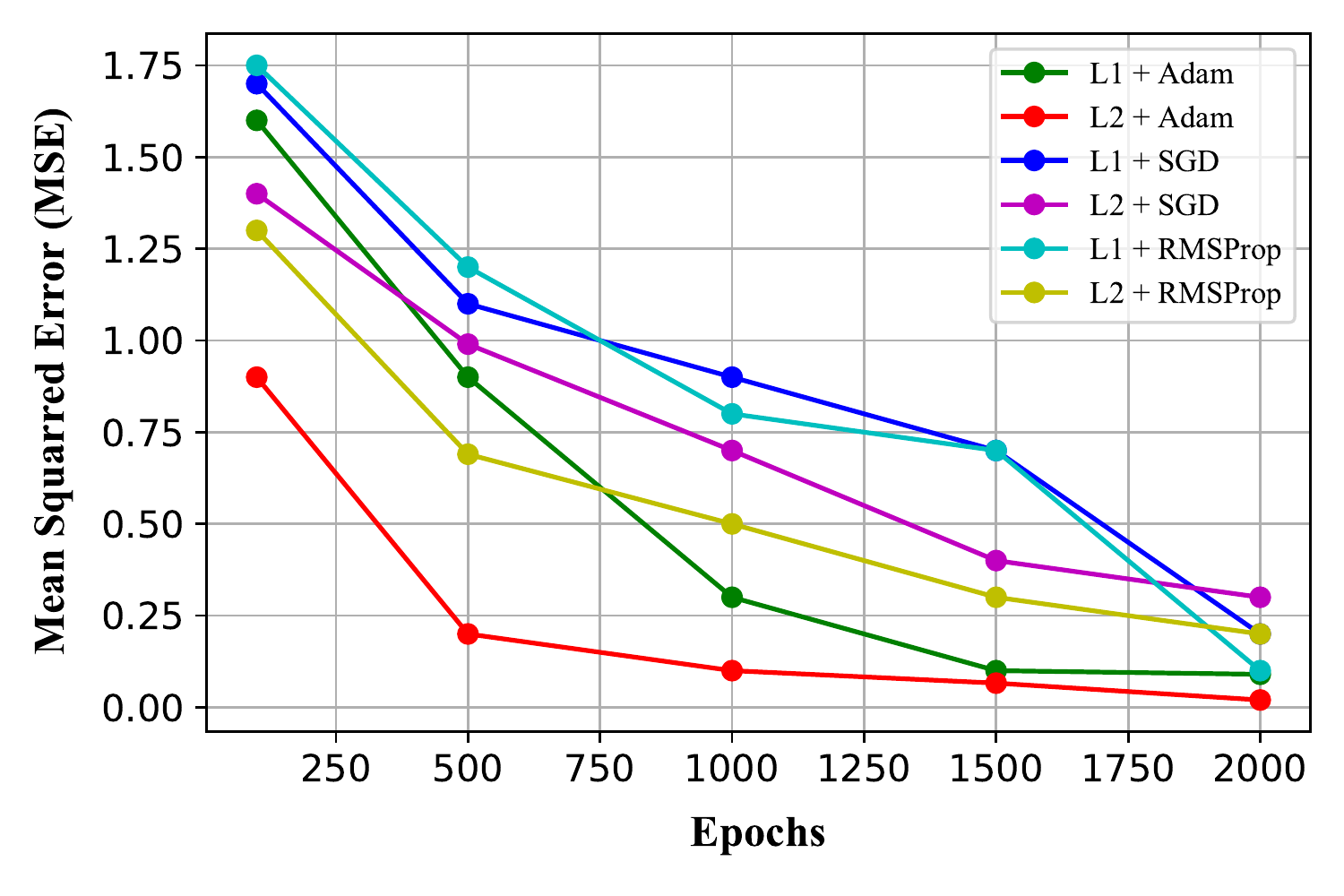}
    }
    \hfill
    \subcaptionbox{Various feature learning units with respect to epochs.}{
     \includegraphics[width = 0.45\linewidth]{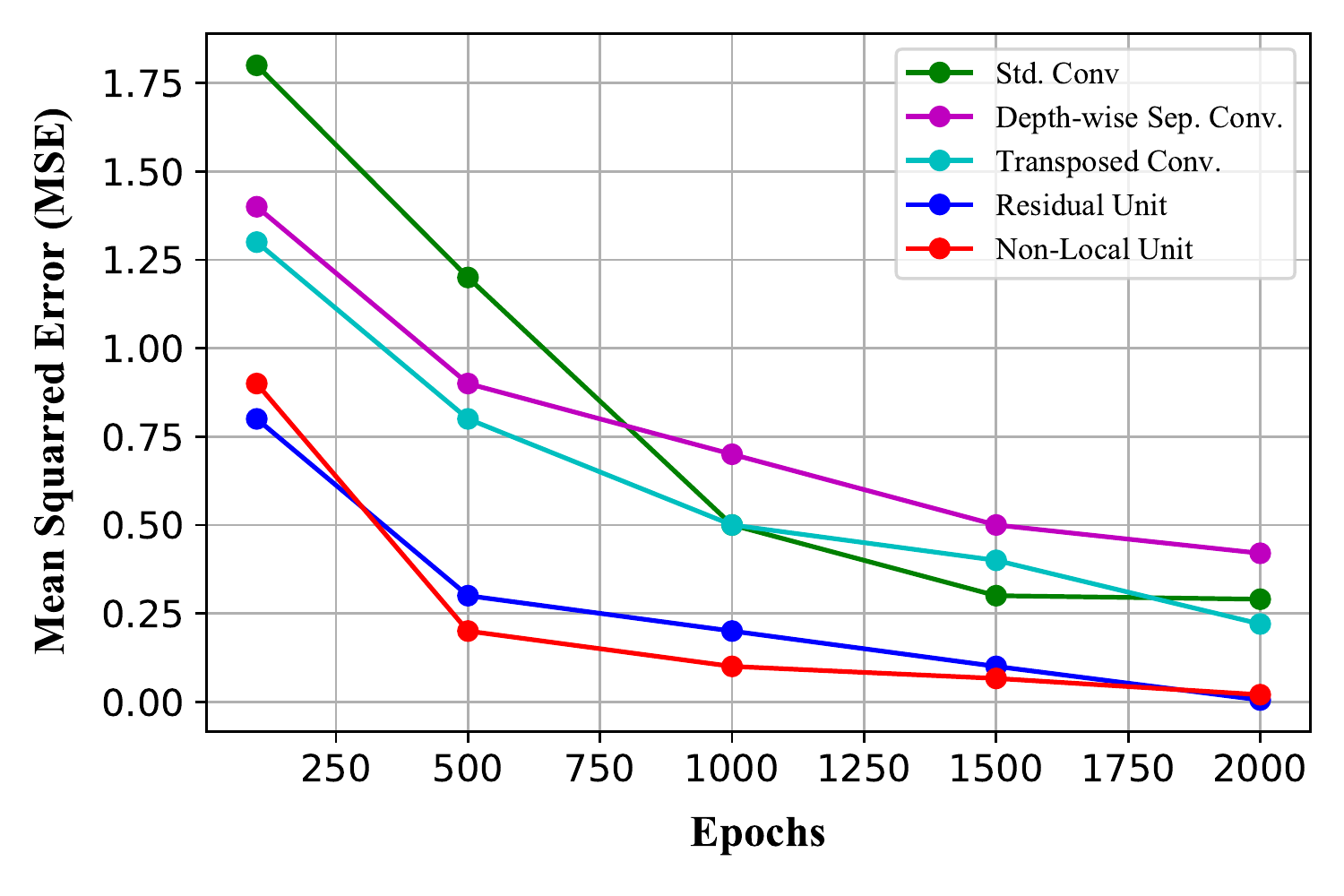}
    }
    }
    \caption{Ablation study: Evaluation of loss functions and feature learning blocks on the Set5 dataset}
    \label{fig:ablation}
\end{figure*}

\subsubsection{Quantitative} Table~\ref{tab:performance} reports the quantitative results for ($\times3$), ($\times4$), ($\times8$) SR methods. Both learning-based (Bicubic, A+, Super-Resolution CNN (SRCNN), Fast SRCNN (FSRCNN), Very Deep Super-Resolutio (VDSR), Laplacian Pyramid Super-Resolution Network (LapSRN), Memory Network (MemNet) and reconstruction-based methods (SRGAN, ESRGAN) have been compared against the proposed framework. It is worth mentioning that deeper architectures perform better over the shallower networks. We also note that larger scaling factors affect the performance of the existing external methods. Among learning-based methods, MemNet demonstrates good performance over large scaling factors because of its large architecture, but its performance drops when the scaling factor is relatively small. The reconstruction-based strategies have higher structural similarity than other methods. Most importantly, our proposed \texttt{NLVAE} model outperforms other reconstruction-based methods in all the scaling factors, generating high-resolution photo-realistic images. This justifies the incorporation of the non-local convolutional block which enables the model to perform better, specifically, on smaller scaled images. Moreover, the deeper architecture of the generative models enhances the performance on large scaling factors, leading the network to a robust zero-shot super-resolution network.

\subsubsection{Qualitative} Fig.~\ref{fig:learning} and Fig.~\ref{fig:reconstruction} depict the visualization of learning-based methods and reconstruction-based methods respectively. Samples from the Set14 and General100 datasets have been used in order to visualize the learning-based solutions. It is confirmed from Fig.~\ref{fig:learning} that our solution produces sharp edges and avoids any undesirable artifacts. As can be seen, a featured generative solution learns better representations between an LR image \& its corresponding HR image. For a fair comparison with reconstruction-based solutions, we utilized Set5, Urban100, and BSDS100 for qualitative comparison among generative SISR models. The visual property of the reconstructed images is magnificent compared to other methods because of its global contextual feature learning process. Fig.~\ref{fig:reconstruction} demonstrates that our method can reduce the blurring artifacts presenting a powerful feature learning ability. It is imperative that the \texttt{NLVAE} model provides better details in regions of irregular structures. More detailed visualizations containing random HR samples from the generated sets and real datasets are provided in the supplementary material.

\section{Ablation Study}

\subsection{Loss function \& Optimizers} In Fig.~\ref{fig:ablation}(a), we have explored different loss functions and optimizers to evaluate the performance of our proposed model. We observe that the combination $L2$ + \texttt{Adam} converges to gradient more smoothly than any other solutions. Among all optimizers, \texttt{Adam} converges to gradient faster than others. \texttt{SGD} and \texttt{RMSProp} provide competitive results but developing slower solutions for a zero-shot process. Between $L2$ and $L1$ loss functions, $L2$ provides faster training and finer reconstruction quality. All the hyperparamters are fixed for this ablation study.

\subsection{Feature Extraction Blocks} To verify the robustness of our non-local convolutional block, we explored various feature extraction units. Fig.~\ref{fig:ablation}(b) shows that non-local convolutional unit performs better than other feature learning units. We observe that the residual unit learns slightly better representations than other units but has relatively larger computational burdens. Comparing our non-local unit against other traditional convolution operations (including depth-wise separable convolution, transposed convolution, and standard convolution operations) our method shows excellent performance with the lowest MSE between LR \& HR images.

\subsection{Feature Extraction Blocks} To verify the robustness of our non-local convolutional block, we explored various feature extraction units. Fig.~\ref{fig:ablation}(b) shows that non-local convolutional unit performs better than other feature learning units. We observe that the residual unit learns slightly better representations than other units but has relatively larger computational burdens. Comparing our non-local unit against other traditional convolution operations (including depth-wise separable convolution, transposed convolution, and standard convolution operations) our method shows excellent performance with the lowest MSE between LR \& HR images.

\subsection{Non-Local Blocks} In this ablations, we study the essentially of non-local blocks for image generation. In table \ref{tab:non_local}, the PSNR and SSIM values are depicted against number of non-local blocks for our proposed method. We note that an increase in non-local blocks provide more accurate image but also increases computational resources. Moreover, we note unusual instability using 5 or more non-local blocks with \textit{ADAM} optimizer. 

\begin{table}[t!]
\setlength{\tabcolsep}{3pt}
\centering
\caption{Number of Non-local blocks for convolutional encoder \& decoder on Set5 dataset}
\medskip
\label{tab:non_local}
\resizebox{0.95\linewidth}{!}{
\begin{tabular}{l|l|l|l}

Convolutional Encoder & PSNR  & Convolutional Decoder & PSNR  \\ \midrule
Non-Local Block - 1   & Unstable & Non-Local Block - 5   & 31.83 \\ \midrule
Non-Local Block - 2   & 27.45 & Non-Local Block - 6   & 32.29 \\ \midrule
Non-Local Block - 3   & 30.12 & Non-Local Block - 7   & 33.81 \\ \midrule
Non-Local Block - 4   & 33.27 & Non-Local Block - 8   & 33.97 \\ \midrule
Non-Local Block - 5   & 34.10 & Non-Local Block - 9   & 34.10 \\ \bottomrule

\end{tabular}
}
\end{table}

\section{Discussions}
In this subsection, we discuss the similarity, dissimilarities and limitations of our method compared to other data-driven strategies. Table \ref{tab:discussion} shows that the input image is linearly upscaled before processing for super-resolution. Similar to  VDSR and DRCN, we also upscale the low-resolution image but linearly. The reconstruction process in our method is progressive as we combine both learning-based and reconstruction-based methods. Learning-based methods generally utilized direct reconstruction of HR images. We adopt the $L2$ loss function for faster convergence in order to maintain high-reconstruction quality. As mentioned above, we do not use a residual representation learning process due to the computational cost for self-supervised settings. Our settings use small modifications of self-supervised settings. We do not use batches of images per epochs; instead we utilize fake batches of a single image for every epochs. Moreover, we perform all these experiments on different sizes of dataset to explore structural variation. Experiments have done on small datasets (Set5, Set14) as well as large datasets (Manga109, Urban100, BSD100) to justify the performance of our proposed solution.

\begin{table}[ht!]
\setlength{\tabcolsep}{6pt}
\centering
\caption{A comparison among various SISR methods defining the loss function, input types, reconstruction types and feature extraction modules.}
\medskip
\label{tab:discussion}
\resizebox{0.98\linewidth}{!}{
\begin{tabular}{c|c|c|c|c}
\toprule
Methods & \begin{tabular}[c]{@{}c@{}}Residual\\ Features\end{tabular} & \begin{tabular}[c]{@{}c@{}}Input\\ Types\end{tabular} & \begin{tabular}[c]{@{}c@{}}Reconstruction \\ Types\end{tabular} & \begin{tabular}[c]{@{}c@{}}Loss\\ Function\end{tabular} \\  \bottomrule \midrule
SRCNN   & No                                                          & LR                                                    & Direct                                                          & L2                                                      \\ \midrule
FSRCNN  & No                                                          & LR                                                    & Direct                                                          & L2                                                      \\ \midrule
VDSR    & Yes                                                         & LR + Bicubic                                          & Direct                                                          & L2                                                      \\ \midrule
DRCN    & Yes                                                         & LR + Bicubic                                           & Direct                                                          & L2                                                      \\ \midrule
LapSRN  & Yes                                                         & LR                                                    & Progressive                                                     & L1                                                      \\ \midrule
\textbf{NLVAE}  & No                                                         & LR + Linear                                                  & Progressive                                                     & L2                                                      \\ \bottomrule
\end{tabular}
}
\end{table}

\section{Conclusions}
We have presented \texttt{NLVAE}, an untrained generative model, featuring a neural encoder-decoder framework capable of reconstructing high-resolution images. With the use of non-local convolutional modules, the model is enabled to capture high-quality semantic information. In addition, the beta variational autoencoder provides more disentangled information reconstructing high-resolution images. Combining the learning-based and reconstruction-based methods, the present method generates sharp and photo-realistic images. The effectiveness of the present model has been confirmed through an extensive experimentation compared with a number of SOTA methods, both qualitatively and quantitatively on multiple benchmark datasets. Moreover, leveraging the power of robust feature learning and generative modeling, the proposed model obviates the need for a large scale dataset while performing SISR. It is to be noted that our proposed method relies on linear upsampling before the super-resolution task. Our future work will include further validation of the NLVAE model against more challenging data settings across various domains as well as more powerful automatic upsampling strategy. We envision more extensive comprehension of our model and more intuitive design of the objective function .

\bibliographystyle{IEEEtran}
\bibliography{document}

\end{document}